\documentclass[12pt,preprint]{aastex}

\begin{document}

\title{Chandra Observations of the Central Region of Abell 3112}

\author{Motokazu Takizawa\altaffilmark{1,2},
        Craig L. Sarazin\altaffilmark{1}, 
        Elizabeth L. Blanton\altaffilmark{1,3},
        and Greg B. Taylor\altaffilmark{4}}

\altaffiltext{1}{Department of Astronomy, University of Virginia, 
                 P. O. Box 3818, Charlottesville, VA 22903; 
                 mt9r@virginia.edu, cls7i@virginia.edu, elb6n@virginia.edu}
\altaffiltext{2}{Department of Physics, Yamagata University, Yamagata, 
                 990-8560, Japan; takizawa@sci.kj.yamagata-u.ac.jp} 
\altaffiltext{3}{{\it Chandra} Fellow}
\altaffiltext{4}{National Radio Astronomy Observatory, P. O. Box O, 
                 1003 Lopezville Road, Socorro, NM 87801; gtaylor@nrao.edu}

\begin{abstract}
We present the results of a {\it Chandra} observation of the central region
of Abell 3112. This cluster has a powerful radio source in the center
and was believed to have a strong cooling flow. The X-ray image shows
that the intracluster medium (ICM) is distributed smoothly on large scales,
but has significant deviations from a simple concentric elliptical
isophotal model near the center.
Regions of excess emission appear to surround two lobe-like radio-emitting
regions.  This structure probably indicates that hot X-ray gas and
radio lobes are interacting.
From an analysis of the X-ray spectra in annuli, we found clear evidence
for a temperature decrease and abundance increase toward the center. The
X-ray spectrum of the central region is consistent with a single-temperature
thermal plasma model.  The contribution of X-ray emission from a
multiphase cooling flow component with gas cooling to very low temperatures
locally is limited to less than 10\% of the total emission.
However, the whole cluster spectrum indicates that the ICM is cooling
significantly as a whole, but in only a limited temperature range
($\geq 2$ keV).
Inside the cooling radius, the conduction timescales based on the Spitzer
conductivity are shorter than the cooling timescales. 
We detect an X-ray point source in the cluster center which is coincident
with the optical nucleus of the central cD galaxy and the core of the
associated radio source.  The X-ray spectrum of the central point source
can be fit by a 1.3 keV thermal plasma and a power-law component whose
photon index is 1.9.  
The thermal component is probably plasma associated with the cD galaxy.
We attribute the power-law component to the central AGN.
\end{abstract}


\section{Introduction}
The central regions of clusters of galaxies are very interesting and 
enigmatic places.
The intracluster medium (ICM) in the central region is often so dense 
that its radiative cooling time is significantly shorter than the Hubble 
time.
Therefore, the ICM will cool down unless it is heated significantly.
Unbalanced radiative cooling would
cause a ``cooling flow'' in the cluster center (see Fabian 1994 for a
review). X-ray imaging analyses 
by {\it ROSAT} and {\it EXOSAT} indicated that mass deposition rates,
$\dot{M}$, were more
than $100 \, M_{\odot} \, {\rm yr}^{-1}$ for many clusters
(Edge, Stewart, \& Fabian 1992; Allen \& Fabian 1997;
Peres et al.\ 1998). 
However, the fate of the resulting cold gas, or 
whether ICM in ``cooling flow'' 
clusters really cools down to low temperatures, is still unclear.
The cD galaxies
in cooling flows have an excess blue star component, which suggests that
the cooled ICM forms new stars (McNamara \& O'Connell 1989). 
However, the implied star formation rates are much lower than the cooling 
flow mass deposition ones (McNamara \& O'Connell 1989).
Furthermore, recent 
{\it XMM-Newton} high resolution spectroscopic observations have imposed
a strong constraint on the amount of gas cooling down to low temperatures
(Peterson et al.\ 2001; Kaastra et al.\ 2001; Tamura et al.\ 2001),
which is consistent with the fact that the mass deposition rates determined
spectroscopically by {\it ASCA} tended to be lower than those determined by  
{\it EXOSAT} and {\it ROSAT} (see Makishima et al.\ 2001 for a review). 

The cD galaxies in the cooling flow clusters usually host strong 
radio sources, with a central radio core and/or radio lobes.
Observations utilizing {\it Chandra}'s excellent spatial resolution have
revealed interactions between these radio lobes and ICM, which may
affect the dynamical and thermal history of ICM in the cluster center. 
In Hydra A (McNamara et al.\ 2000; David et al.\ 2001), Perseus 
(Fabian et al.\ 2000), and Abell 2052 (Blanton et al.\ 2001), radio lobes 
coincide with X-ray low surface brightness regions, which suggests that
the radio plasma expands to displace the ICM. 
On the other hand, X-ray cavities 
without radio emission are also found at larger distances 
in Abell 2597 (McNamara et al.\ 2001). 
They are believed to be old radio 
lobes which cannot emit (at least at high radio frequencies) because the high 
energy relativistic electrons have already suffered radiative losses.
These radio quiet cavities also 
indicate that cD galaxies provided the ICM with high energy electrons and 
magnetic fields intermittently.

Abell 3112 is a cooling flow cluster at a redshift of $z=0.0746$. There 
is a powerful radio source, PKS 0316-444, in the cluster center.
Previous X-ray imaging analyses with {\it EXOSAT}
(Edge, Stewart, \& Fabian 1992)
and {\it ROSAT}
(Allen \& Fabian 1997; Peres et al.\ 1998) indicated that Abell 3112 had
a strong cooling flow, with a
mass deposition rate of $\dot{M} \approx 400$ $M_{\odot}$ yr$^{-1}$.
The temperature profile at a larger scale than the cooling radius was studied
by Markevitch et al.\ (1998) and by Irwin, Bregman, \& Evrard (1999) with 
{\it ASCA} data. Markevitch et al.\ (1998) found that the temperature 
decreased outward from 6 keV to 3 keV with {\it ASCA} data, although they 
explicitly assumed a cooling flow component in the central region. 
On the other hand, Irwin et al.\ (1999) argued that the data 
were consistent with an isothermal distribution.
An increase in the iron abundance toward 
the center was reported by Finoguenov, David, \& Ponman (2000). 
However, there was no detailed study of the temperature and abundance
structure within the cooling radius ($\la 250$ kpc) because of the limited
spatial resolution of {\it ASCA}.
Furthermore, any interaction region between the radio source and the ICM
would be too small to have been resolved with {\it ROSAT} or {\it ASCA}.
In this paper, we present {\it Chandra} observations of central region of
Abell~3112.
We assume $H_0 = 50$ km s$^{-1}$ Mpc$^{-1}$, $\Omega_0 = 1.0$, and 
$\Omega_{\Lambda} = 0.0$.
At a redshift $z=0.0746$, 1\arcsec\ corresponds to
1.94 kpc. 
The errors correspond to the 90\% confidence level throughout the paper.

\section{Observations and Data Reduction}

\subsection{X-ray Observations}

Abell 3112 was observed twice with {\it Chandra} on 2001 May 24 for
7,257 seconds and on 2001 September 15 for 17,496 seconds.
Each observation was pointed so that the cluster center would fall near
the aimpoint of the ACIS-S3 detector.
However, the roll angles were different, so the outer parts of the field
of view differ for the two observations.
The two observations were merged, using the positions of bright point
sources to register the two images.
Here, we analyze data from the S3 chip only.
The data from the S1 chip was used to search for the background flares.
A few short periods with background flares were found and removed
using the {\sc lc\_clean} software provided by
Maxim Markevitch\footnote{See
\url{http://asc.harvard.edu/cal/Links/Acis/acis/Cal.prods/bkgrnd/current/index.html}.},
leaving a total exposure of 21,723 seconds.
Only events with ASCA grades\footnote{See \url{http://asc.harvard.edu/ciao/}.}
of 0, 2, 3, 4, and 6 were included.
Standard bad pixels and columns were removed.
Exposure maps and background data were generated for each observation
separately, and then merged appropriately.
Blank sky background data were taken from the compilation by
Maxim Markevitch\footnotemark[5].

The data are possibly affected by the low energy QE degradation of 
ACIS.\footnote{See 
\url{http://cxc.harvard.edu/cal/Links/Acis/acis/Cal\_prods/qeDeg/index.html}
.}
We tried to correct this using the {\sc corrarf}\footnotemark[7] program.
However, the correction did not appear to work well for this data;
specifically, {\sc corrarf} seemed to over-correct the data.
As a result,
we obtained absorption column densities which were much lower than 
the Galactic value in our spectral fits.
We expect that the absorbing column should be at least as large 
as the Galactic value.
Therefore, 
we will show the results without the correction for the low energy QE degradation.
Fortunately, this correction significantly affects only the very low energy 
($\le$ 1 keV) part of the {\it Chandra} band.
As a result, neither temperature nor abundance 
determinations were seriously affected by this correction. 
On the other hand, mass deposition rates of cooling flow components
increased by a factor of about two when we fitted the spectra 
after the correction.

\subsection{Radio observations}

The observations were made with Very Large Array\footnote{The National
Radio Astronomy Observatory is operated by Associated Universities,
Inc., under cooperative agreement with the National Science
Foundation.} at a center frequency of 1320 MHz on 1996 October 18.
The array was in the `A' configuration, which provided an angular
resolution of 6.9 $\times$ 1.5 arcsec.  A total of 1.2 hours were obtained
on source using 32 channels across a 6.25 MHz band.  Both right and
left circular polarizations were observed.  Phase and bandpass
calibration was obtained by short (1 min) observations of the strong
(4.9 Jy) calibrator J0440{\tt -}4333 taken every $\sim$20 minutes.
Atomic hydrogen absorption was searched for towards the compact core
over a velocity range from 22,100--23,440 km s$^{-1}$ at a resolution of
44 km s$^{-1}$, but not detected down to a 3$\sigma$ limit
on the maximum optical depth of 0.012.

\section{X-ray Images}

\begin{figure}
\epsscale{0.8}
\plotone{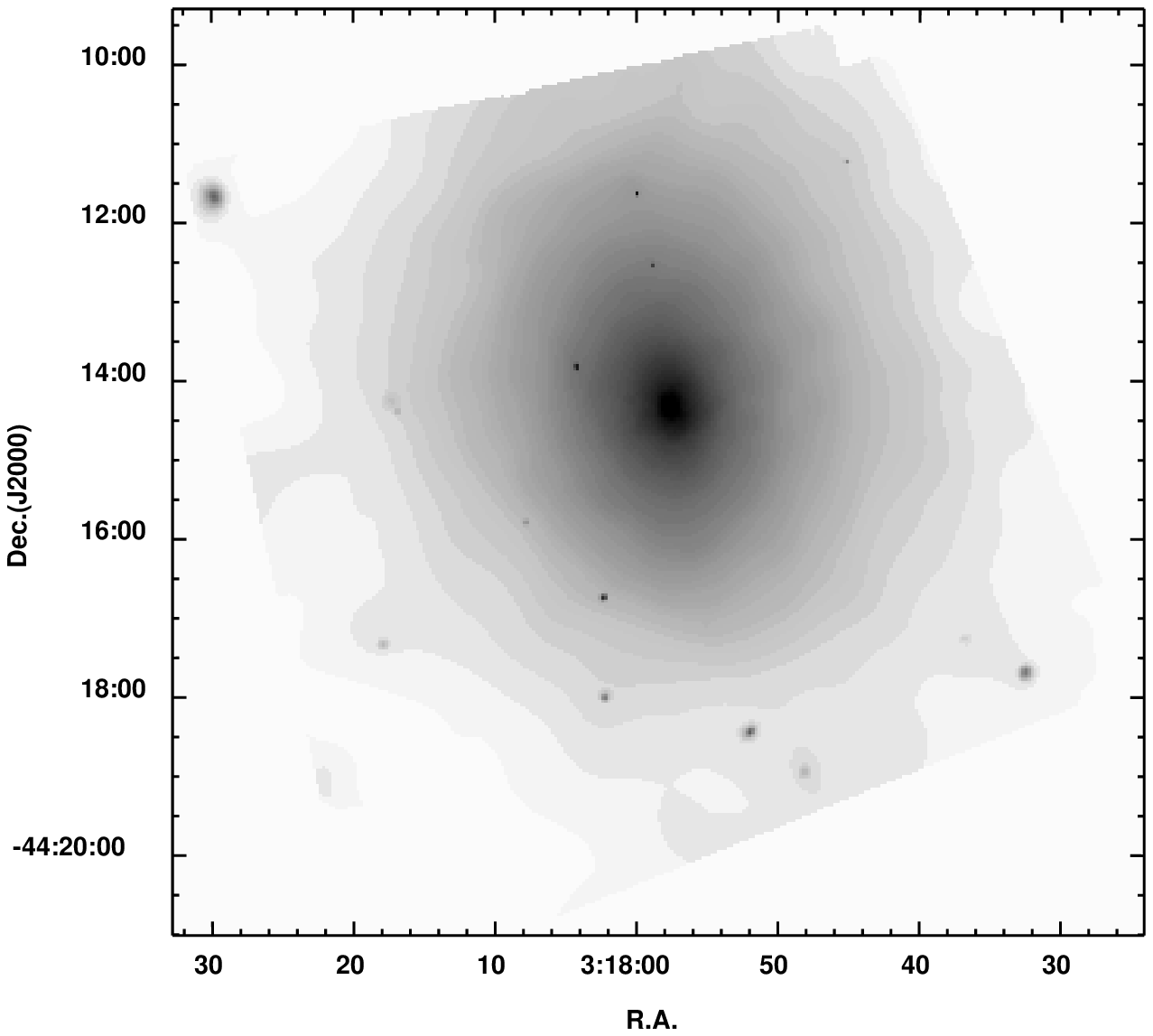}
\caption
{Adaptively smoothed {\it Chandra} image (0.3-10.0 keV) with a minimum
signal-to-noise of 3 per smoothing beam.
The image has been background subtracted and corrected for exposure and
vignetting.
 \label{fig:corimage}}
\end{figure}

The X-ray image was adaptively smoothed using the
{\sc ciao}\footnote{See \url{http://asc.harvard.edu/ciao/}.} {\sc csmooth}
program with a minimum signal-to-noise of 3 per smoothing beam.
An identically-smoothed background was subtracted, and the image
was divided by an identically-smoothed exposure map.
The exposure map was corrected for vignetting,
using the typical energy of the observed 
photons to take account of the energy dependence of the vignetting.
The resulting image is shown in Figure \ref{fig:corimage}.
As in other typical cooling flow clusters,
X-ray emission is distributed symmetrically and is fairly concentrated in
the center.
The image is smooth and quite symmetric on large scales.
The cluster emission is elliptical, with a
major axis along the direction from NNE to SSW.
On large scales, there appears to be no substructure
except for some point sources.

We used a wavelet detection algorithm (WAVEDETECT in CIAO) to detect
individual point sources. Sources were visually confirmed on the X-ray image,
and a few low-level detections were removed. Using this method, we found
sixteen point sources. One of these corresponds to the radio source
PKS 0316-444, which is located at the center of the central
cD galaxy ESO 248- G 006.
Another corresponds to a galaxy in the cluster (LCRS B031619.9-442441). 

The regularity of the cluster X-ray emission on large scales
doesn't continue into the very central region. 
Figure \ref{fig:center1}
shows the X-ray contour map of the $\sim 1' \times 1'$ region around
the central point source. In the very center, an elongated structure 
from the central point source towards the
SE is clearly seen.
We can see another filament-like component from $\sim 10\arcsec$ south of 
the central point source towards the SW.
To clarify structures
which deviate from symmetric cluster emission, 
we fitted the entire cluster image
with a concentric elliptical isophotal model, and then subtracted 
the best-fit model
from the image to get the residual component.
In the fitting procedure, we fixed the center of the isophotes to
the center determined from the larger scale isophotes about which the image was
symmetric, rather than the central point source.
The ellipticity and position angle for each isophote were allowed to vary.
It is most likely that relaxed cluster potential structure is not spheroidal
but triaxial (Jing \& Suto 2001).
In this case, the ICM hydrostatic density distribution also will be triaxial
(Lee \& Suto 2003).
Therefore, the ellipticities and the position 
angles of X-ray isophotes can vary with radius unless our line-of-sight
coincides with one of the three symmetry axes. 

In Figure~\ref{fig:center2}, the greyscale shows the residual produced by
subtracting our best-fit elliptical isophotal model from the adaptively
smoothed image of the central region of the cluster.
The sharp elliptical edge near the outside is an artifact because
the elliptical model extended only to this isophote.
Dark areas are positive residuals (excess emission), while light
areas are negative residuals.
The dark and light regions at the center are due to the central point
source, which is not exactly at the center of the X-ray isophotes at
large  radii.
There are regions of excess emission to the south of the central source.
The contours show the 1.32 GHz VLA radio image.
There is a very bright radio core, which is unresolved in the radio image;
the elliptical contours for the core show the beam of the radio
observation.
The very bright radio core is coincident with the central X-ray source.
There are two diffuse radio regions to the SE and SW of the radio core,
which may be connected to the core.
The NS extent of these regions is presumably exaggerated due to the
elongated observing beam.
Roughly speaking, the excess X-ray regions appear to
surround two lobe-like radio components.
However, part of the SW radio lobe is coincident with a region of 
X-ray excess emission.
Unfortunately, the low resolution of the radio data, bright radio core, and
elongated radio beam make it difficult
to decide definitely whether the excess is in the radio lobes or partly 
surrounding it.

\section{Spectral Analysis} \label{sec:spectrum}

In order to examine the temperature and abundance structure quantitatively,
we determined the X-ray spectra in annuli.
We fitted the data between 0.5 and 10.0 keV with a photo-absorbed single 
temperature MEKAL model (Kaastra 1992; Liedahl, Osterheld, \& Goldstein 1995)
after masking the point sources.
We tried to fit the data fixing the absorbing column density to the Galactic
value ($N_H = 2.51 \times 10^{20}$ cm$^{-2}$; Dickey \& Lockman 1990) 
or allowing it to vary freely.
The fitting results depend somewhat on whether we fixed the absorption 
to the Galactic value or not. 
Figure~\ref{fig:radtemabund} shows the radial 
temperature and abundance profiles.
The solid crosses are the values obtained 
when we allowed the absorption to vary,
while the dashed crosses are the values when we fixed the absorption to
the Galactic value. 
The dashed crosses are shifted slightly to the left
in order to be more easily seen.
The results with fixed or varying absorption are similar.
At $r > 70\arcsec$, both the temperature and abundance are nearly 
constant at $k T \sim 6$ keV and $Z \sim 0.4 \, Z_\odot$, respectively.
The temperature decreases from $\sim 6$ keV at 70\arcsec\ to 3.5 keV
in the central 20\arcsec.
The abundance increases from $\sim 0.4 \, Z_\odot$ at 70\arcsec\ to
1.3  $Z_\odot$ in the central 20\arcsec.

The temperature values obtained with variable absorption, along with
their associated errors, are fitted to a power-law profile given by
$T(r) = T_0 (r/1\arcsec )^p$. 
For $r<70\arcsec$, $T_0 = 1.838 \pm 0.254$ keV and $p = 0.270 \pm 0.044$.
For $r>70\arcsec$, $T_0 = 4.226 \pm 3.174$ keV and $p = 0.064 \pm 0.165$.
We will use these fitted temperature profiles later when we analyze
conduction timescale and integrated gravitational mass.

To examine the non-radial temperature structure, we made a two-dimensional 
temperature map of the central $240\arcsec \times 240\arcsec$ region.
We divided the region into 64 ($8 \times 8$)
square regions, each of which is $30\arcsec \times 30\arcsec$. Then, we fitted 
the data of each region with the photo-absorbed MEKAL model where
absorption is treated as a free parameter. Figure \ref{fig:tmap} shows
a two-dimensional temperature map of the central $240\arcsec \times 240\arcsec$
region overlaid with X-ray surface brightness contours. Black and white 
represent lower ($\sim 3$ keV) and higher ($\sim 6$ keV) temperatures,
respectively.
There is no 
significant azimuthal structure in the temperature map while
temperature decrease toward the center is again clearly seen. 
We also made a two-dimensional abundance map in the same way.
However, abundances in the individual fits were too poorly constrained 
to provide a useful map.

We tried to fit the annular spectra with a model for a multiphase
cooling flow to constrain the contribution from a cooling flow component 
and a mass deposition rate in each annulus, although the data within each
annulus are almost consistent with a single temperature MEKAL model. 
As a cooling flow spectral model, we used the
MKCFLOW model based on the MEKAL model. We added a MEKAL model representing 
the emission from the ICM outside the cooling flow. Both MKCFLOW and MEKAL
were assumed to be subject to the same absorption column, which we 
fixed to the Galactic value or let vary freely. We fixed the metallicity
and initial gas temperature of the MKCFLOW model to the same values of 
metallicity and temperature as the MEKAL model, respectively.
The low temperature in the MKCFLOW model is fixed to the lowest value 
(0.001 keV) in XSPEC. 
The results
are shown in Table \ref{tab:cflow1} and \ref{tab:cflow2}. 
Compared with the fitting results without
cooling flows, the results are not improved dramatically.
Figure \ref{fig:radmdot} shows 
a radial profile of the mass deposition rate derived from the fitting.
The circles and solid error bars are the values obtained from fits with
freely varying absorption. The squares and dashed error bars are the values
from the fits with fixed Galactic absorption.
When absorption is set as a free parameter, forth bin and the outermost 
three bins are consistent with a model without a cooling flow component.
Only the inner three bins and the fifth bin have the cooling flow component,
although
their errors are fairly large and comparable to themselves. The resultant 
total mass deposition rate is $44.5^{+52.1}_{-32.5} \, M_{\odot}$ yr$^{-1}$,
which is significantly smaller 
than the rates derived from the previous image analyses of {\it EXOSAT} and 
{\it ROSAT} data, 
which are $\sim 400 \, M_{\odot}$ yr$^{-1}$
(Edge, Stewart, \& Fabian 1992; Allen \& Fabian 1997;
Peres, et al. 1998).
The cooling flow component contributes less than
$\sim 10$\% of the total X-ray emission. When we adopt 
the Galactic absorption value, the mass deposition rate becomes even lower
and consistent with a model without a cooling flow component in all bins.

Even though multiphase gas is not found within small regions of the cluster,
we do see an overall temperature gradient, with gas in 
the center that is cooler by a factor of $\sim 2$ compared with 
gas in the outer regions of the cluster. Therefore, if we fit a spectrum 
extracted from the whole cluster, a significant mass deposition rate and 
a low temperature cutoff might be expected. 
We fit the data within $157\arcsec$ with the MEKAL plus MKCFLOW model where 
the low temperature in MKCFLOW is allowed to vary.
The results are shown in Table~\ref{tab:totspec}. The upper and lower
rows show the results with the absorption column allowed to vary and fixed
to the Galactic value, respectively. 
In contrast to the spectral analyses in annuli, 
the mass deposition rate is comparable with the former values 
based on imaging analysis. However, the low temperature in the MKCFLOW
model is not very low ($kT_{\rm Low} \sim 2$ keV), which means that the 
observations are consistent with significant cooling, but only 
over a limited temperature range. This is consistent with what has been
observed with many cooling flow clusters using Chandra and XMM-Newton
(e.g., Peterson et al.\ 2003).
Please note that the goodness of the fit is marginal
(the reduced chi-squared is about 1.5).
This might indicate
that the emission measure distribution with the temperature is not
that of the standard cooling flow model even in the temperature range
between $T_{\rm High}$ and $T_{\rm Low}$.

\section{X-ray Profiles and Deprojection}

To quantify the radial structure of the ICM, we made a radial profile 
of the X-ray
surface brightness in the 0.3-10.0 keV band (Figure \ref{fig:radsfpden}a). 
The bins for X-ray surface brightness were chosen to be the same as those
used below to determine the radial variation in the X-ray spectrum
(\S~\ref{sec:spectrum}).
The surface brightness values were deprojected to determine the X-ray
emissivity and gas density (Figure \ref{fig:radsfpden}b), assuming the
emissivity is constant in spherical shells. 
When we convert the emissivity into the gas density, 
we use a MEKAL code and assume that the temperature is constant
in spherical shells and equal to that derived from the spectral 
fitting of projected data. 
Some fluctuations are seen 
in the density and pressure profiles near the outer boundary
(Figure \ref{fig:radsfpden}b and \ref{fig:radsfpden}c), which are
artifacts of the deprojection since we assume zero emissivity outside the 
outer boundary. 
However, many simulations of our deprojection method show 
that this affects only a few outermost points, and has no significant 
effect anywhere near the center.
The observed gas density and pressure profiles are quite smooth outside
of the region of the radio source.
Based on the smooth radial profiles and lack of substructure in the
X-ray image outside of the radio source region, we conclude that this cluster
is dynamically relaxed and that the ICM is generally 
in hydrostatic equilibrium.

\section{Mass Distributions}

The gravitational mass distribution was determined from the equation of 
hydrostatic equilibrium, 
\begin{equation}
  M_{\rm tot}(<r) = - \frac{k_B T r}{\mu m_p G} \biggl(
      \frac{d \ln \rho_{\rm gas}}{d \ln r} + \frac{d \ln T}{d \ln r} 
                  \biggr).
\end{equation}
We use the fitted power-law temperature profiles when $M_{\rm tot}(r)$
is calculated. To reduce the noise 
in the density gradient term, we determine the gradient at $r_i$ 
by differencing the densities at $r_{i-2}$ and $r_{i+2}$. 
Uncertainties in the density gradient at $r_i$ are also calculated 
from the density uncertainties at $r_{i-2}$ and $r_{i+2}$.
The density gradients and their uncertainties
are assumed to be equal in the innermost three points and the outermost 
three points. The gas mass profile is also determined from the ICM
density profile assuming spherical symmetry. The gas mass is then given by
\begin{equation}
  M_{\rm gas}(<r) = \int^r_0 \rho_{gas}(r) 4 \pi r^2 dr
\end{equation}
The integrated gravitational mass and gas mass are shown 
in Figure \ref{fig:radmrmg} (a).

A radial profile of the gas mass fraction is shown 
in Figure \ref{fig:radmrmg} (b).
The gas mass fraction increases from $3\%$ at $10\arcsec$ to $10\%$ at 
$100\arcsec$. This trend is similar to those in other well-relaxed clusters
(David, Jones, \& Forman 1995; Ettori, \& Fabian 1999;
Allen, Schmidt, \& Fabian 2002).
Our gas mass fraction is lower than that of Mohr, Mathiesen, \& Evrard (1999)
at much larger radii ($>$ 1 Mpc), but the gas fractions of clusters
generally increase with radius, and this appears to apply to Abell~3112.
Using the same beta-model fits as in Mohr et al., we find that the gas
fraction at $100\arcsec$ is predicted to be $\sim$12\%, which agrees well
with our values.

\section{Comparison of Cooling and Thermal Conduction}\label{sec:coolcond}

Because the temperature is higher than $\sim 3$ keV over the whole cluster, 
the main emission mechanism is thermal bremsstrahlung. Therefore,
the isobaric cooling time is (Sarazin 1986)
\begin{equation}
  t_{\rm cool} =  8.5 \times 10^{10}
              \biggl( \frac{n_p}{10^{-3} \, {\rm cm}^{-3}} \biggr)^{-1}
              \biggl( \frac{T}{10^8 \, {\rm K}} \biggl)^{1/2} \, {\rm yr} 
\end{equation}
Radiative cooling would make a radial temperature gradient 
in the central region of the ICM through its density dependence,
as is observed.
(Fig.~\ref{fig:radtemabund}).
On the other hand, thermal conduction would have the effect of reducing 
the temperature gradient.
The conduction timescale is generally expressed as (see also Sarazin 1986)
\begin{equation}
  t_{\rm cond} = \frac{n_e l_T^2 k_B}{\kappa}
\end{equation}
where $n_e$ is electron number density, $l_T = T / \nabla T$ is 
the scale length of the temperature gradient, 
and $k_B$ is the Boltzmann constant.
If we consider only the Coulomb scattering process, the thermal conductivity 
for hydrogen plasma is (Spitzer 1962) 
\begin{equation}
\kappa = 4.6 \times 10^{13} 
         \biggl( \frac{T_e}{10^8 \, {\rm K}} \biggr)^{-5/2}
         \biggl( \frac{\ln \Lambda}{40} \biggr)^{-1} \,
         {\rm erg}^{-1} \, {\rm cm}^{-1} \, {\rm K}^{-1} \, ,
\end{equation}
where $\ln \Lambda$, the Coulomb logarithm, is
\begin{equation}
  \ln \Lambda = 37.8 + \ln \biggl[ 
      \biggl( \frac{T_e}{10^8 \, {\rm K}} \biggr)
      \biggl( \frac{n_e}{10^{-3} \, {\rm cm}^{-3} } \biggr)^{-1/2}
                            \biggr].
\end{equation}
Figure \ref{fig:radtctcond} shows radial profiles of $t_{\rm cool}$
and $t_{\rm cond}$. 
When we calculated $t_{\rm cond}$, we used the fitted
power-law temperature profile to reduce the noise in the temperature 
gradient.  Also, $t_{\rm cond}$ is evaluated only for $r < 70\arcsec$ because
an isothermal distribution is consistent with the temperature data
in the outer regions.

The cooling time of the innermost bin is $2 \times 10^9$ yr. If we define 
a cooling radius as where $t_{\rm cool}$ is $1.5 \times 10^{10}$ yr, 
it becomes $\sim 100\arcsec$, or $\sim 200$ kpc. 
Both results are consistent with 
the former {\it ROSAT} (Allen \& Fabian 1997; Peres et al.\ 1998) and 
{\it EXOSAT} results (Edge, Stewart, \& Fabian 1992).
At the innermost three bins, $t_{\rm cool}$ and $t_{\rm cond}$ are comparable 
with each other. Outside of this region, the conduction timescale is clearly
shorter than the cooling timescale. 
If conduction proceeded at the Spitzer rate, it would erase the temperature
gradient.
Since the observations show a significant gradient in this region
(Fig.~\ref{fig:radtemabund}a),
thermal conduction must be significantly suppressed by, e.g., tangled
magnetic fields and/or plasma instabilities.
The cooling rate of gas with solar abundance is only 
$\sim$ 10\% higher than that of hydrogen-helium mixture gas at
3.5 keV (see Figure 9-9 of Binney \& Tremaine 1987). Therefore,
actual cooling timescales in the central regions are probably
$\sim$ 10\% shorter at the most, which does not change 
the situation here dramatically.
Although our estimation here is rough, it is still useful for an estimation
of an order of magnitude. Calculations taking account of time evolution
will be helpful to make more precise models.

\section{The Central Point Source}

We detected an X-ray point source in the cluster center. The position is
coincident with the optical core of the cD galaxy and the radio core.
The position of the X-ray point source is (epoch J2000)
$\alpha=03^h17^m57.65^s$ $\delta = -44^{\circ}14\arcmin17.06\arcsec$ with
an uncertainty of 0.04 arcsec in each coordinate.
The spectrum of the point source is shown in Figure~\ref{fig:cpssp}.
We fitted the data with a MEKAL model plus a power-law model.
In addition to the 
absorption column density common for both components, we added intrinsic 
absorption for the power-law component only. As a result,
the applied model is
\begin{equation}
      {\rm Model}_{\rm AGN} = {\rm WABS} \times 
      [{\rm MEKAL} + {\rm ZABS} \times {\rm POWERLAW}].
\end{equation}
The abundance of the MEKAL component was fixed to the value derived 
from the region surrounding the point source.
The absorbing column $N_{\rm H}$
was also fixed either to the value derived from the surrounding 
region or the Galactic value.
The fitting results are shown in Table~\ref{tab:centps}.
The spectral model consists of a 1.26 keV thermal plasma component and 
a power-law component with a photon index of $\sim 1.9$. 
The power-law component 
has an extra absorbing column of $\sim 2 \times 10^{21}$ cm$^{-2}$.
This is consistent with the central source being a strongly absorbed AGN.

\section{Discussion}

\begin{figure}
\epsscale{0.8}
\plotone{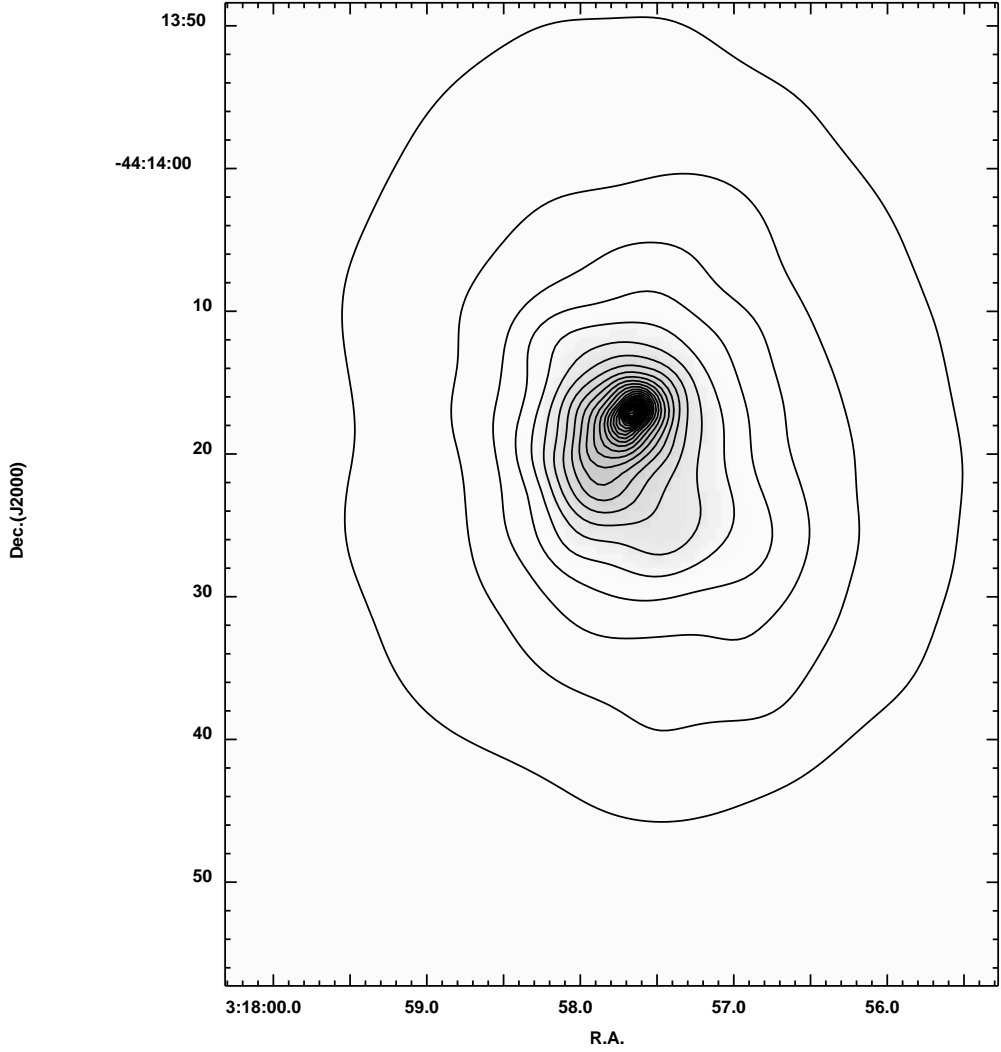}
\caption
{X-ray contour map (0.3-10.0 keV) of the central
$\sim 50\arcsec \times 70\arcsec$ region of the cluster Abell~3112.
Contours are logarithmically spaced.
In the very center, an elongated structure is seen extending
from the central point source towards the SE.
Another filament extends from $\sim 10\arcsec$ south of 
the central point source towards the SW.
 \label{fig:center1}}
\end{figure}

\begin{figure}
\epsscale{0.8}
\plotone{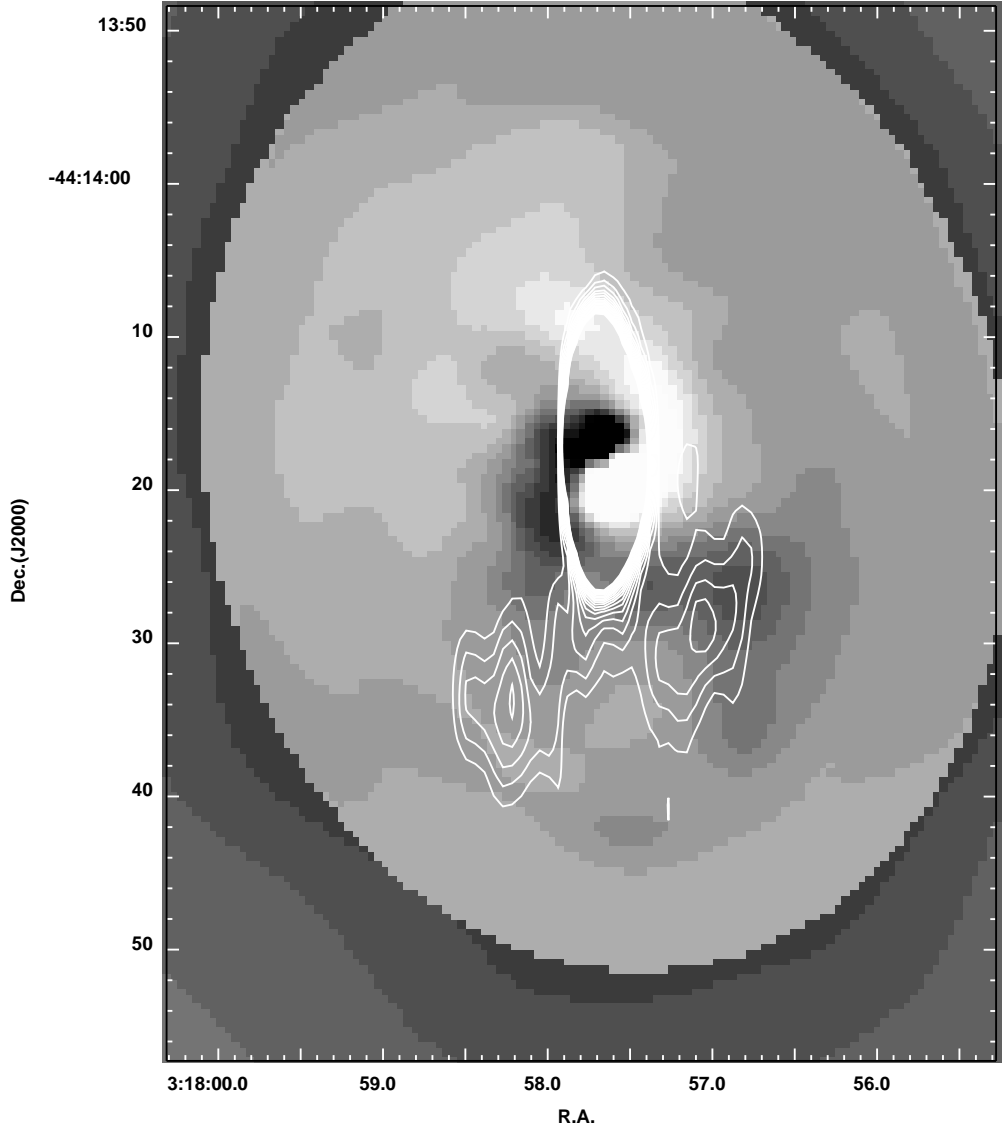}
\caption
{The greyscale shows the residuals of the adaptively-smoothed X-ray
image (0.3--10 keV) from our best-fit elliptical isophotal model.
Dark areas are positive residuals (excess emission), while light
areas are negative residuals.
The dark and light regions at the center are due to the central point
source, which is not exactly at the center of the X-ray isophotes at
large  radii.
The contours are from the 1.32 GHz VLA radio image.
The very bright radio core is coincident with the central X-ray source.
The NS extension of the radio core source is due to the elongated radio
observing beam.
\label{fig:center2}}
\end{figure}

\begin{figure}
\epsscale{0.8}
\plotone{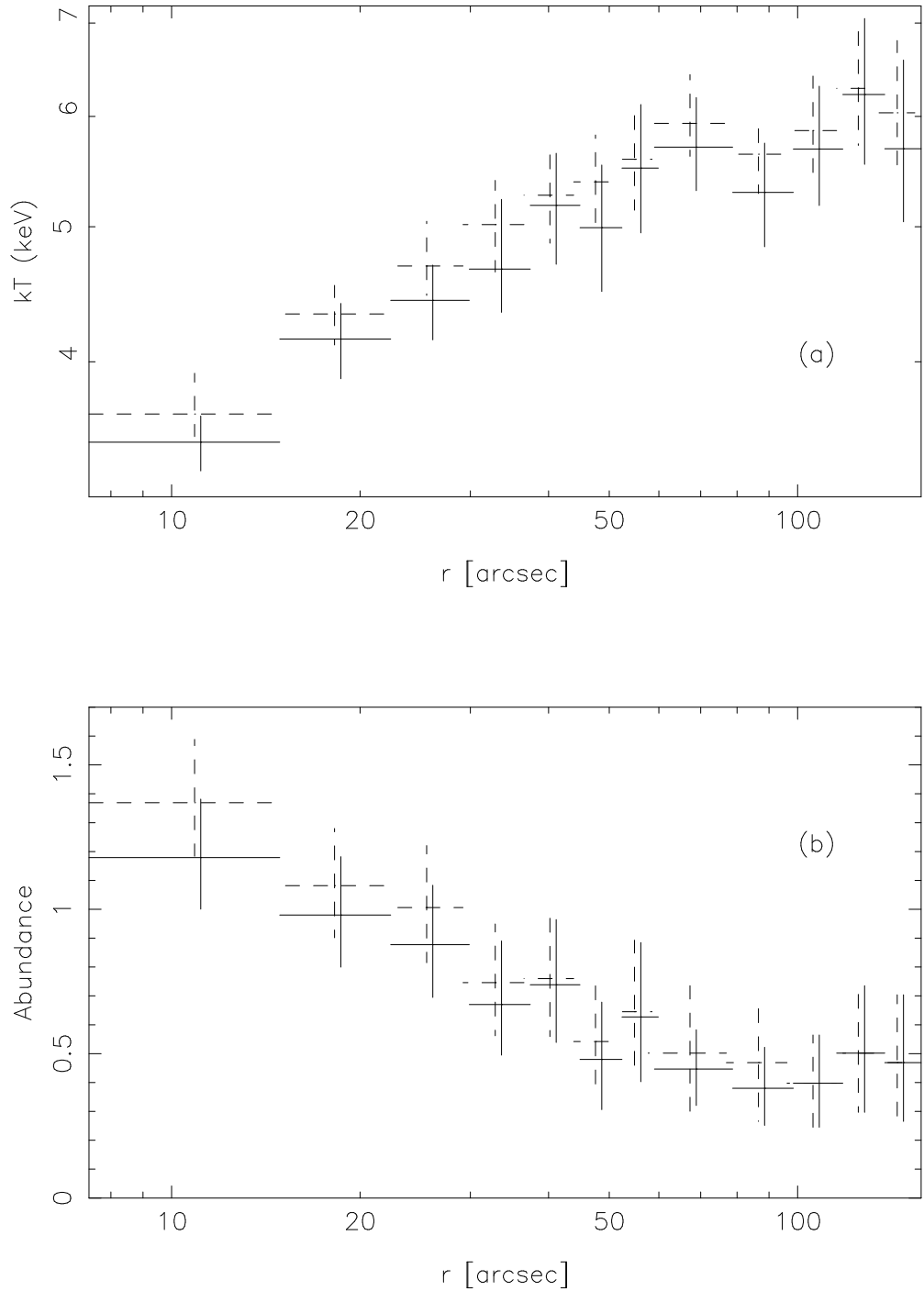}
\caption
{Radial profiles of temperature (a) and abundance (b), respectively.
 The solid crosses are the values obtained when we allowed the absorption
 to vary,  while the dashed crosses are the values 
  when we fixed the absorption to the Galactic value. 
 \label{fig:radtemabund}}
\end{figure}

\begin{figure}
\epsscale{0.8}
\plotone{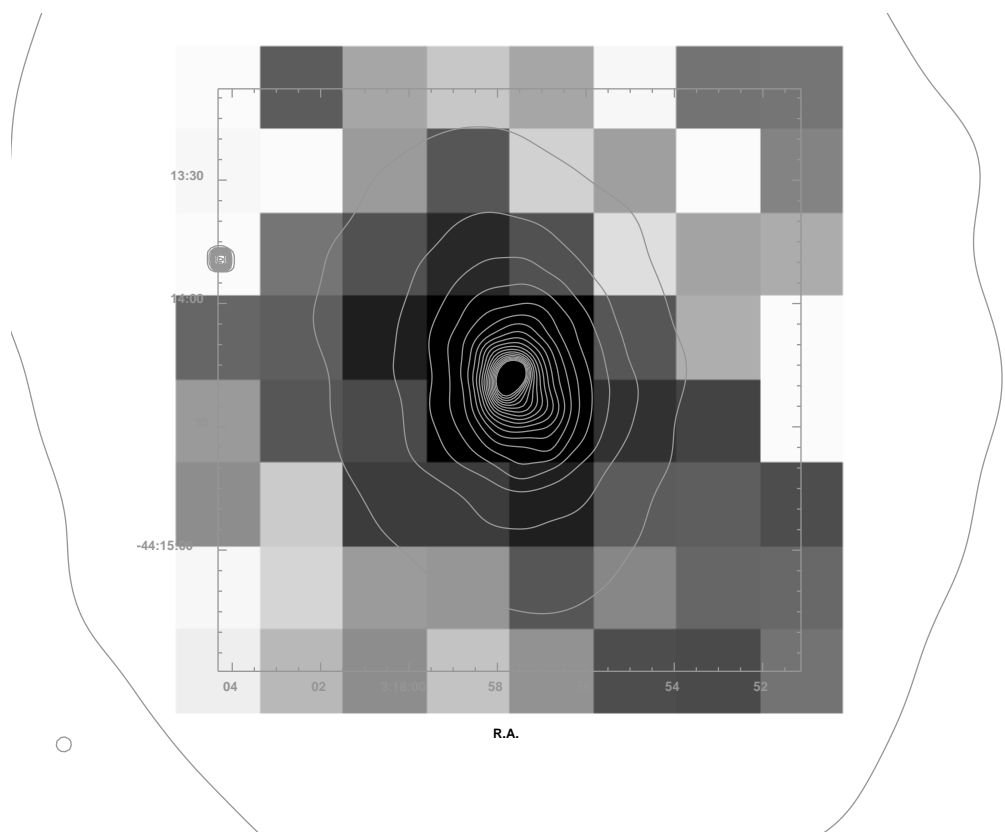}
\caption
{A two-dimensional temperature map of the central
$240\arcsec \times 240\arcsec$ 
 region overlaid with X-ray surface brightness contours. Black and white 
 represent lower ($\sim 3 keV$) and higher ($\sim 6 keV$) temperature, 
 respectively.
\label{fig:tmap}}
\end{figure}

\begin{figure}
\epsscale{0.8}
\plotone{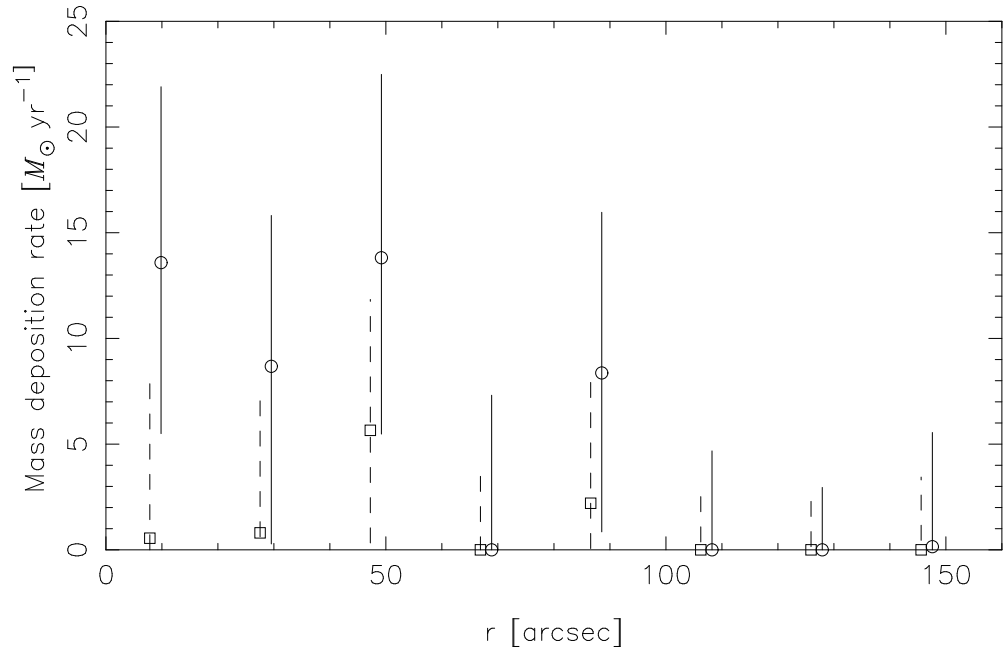}
\caption
{Radial profile of the mass deposition rate.
 The circles and solid error bars are the values obtained 
when we allowed the absorption to vary,  while the squares dashed 
error bars are the values when we fixed the absorption to the Galactic value. 
 \label{fig:radmdot}}
\end{figure}

\begin{figure}
\epsscale{0.8}
\plotone{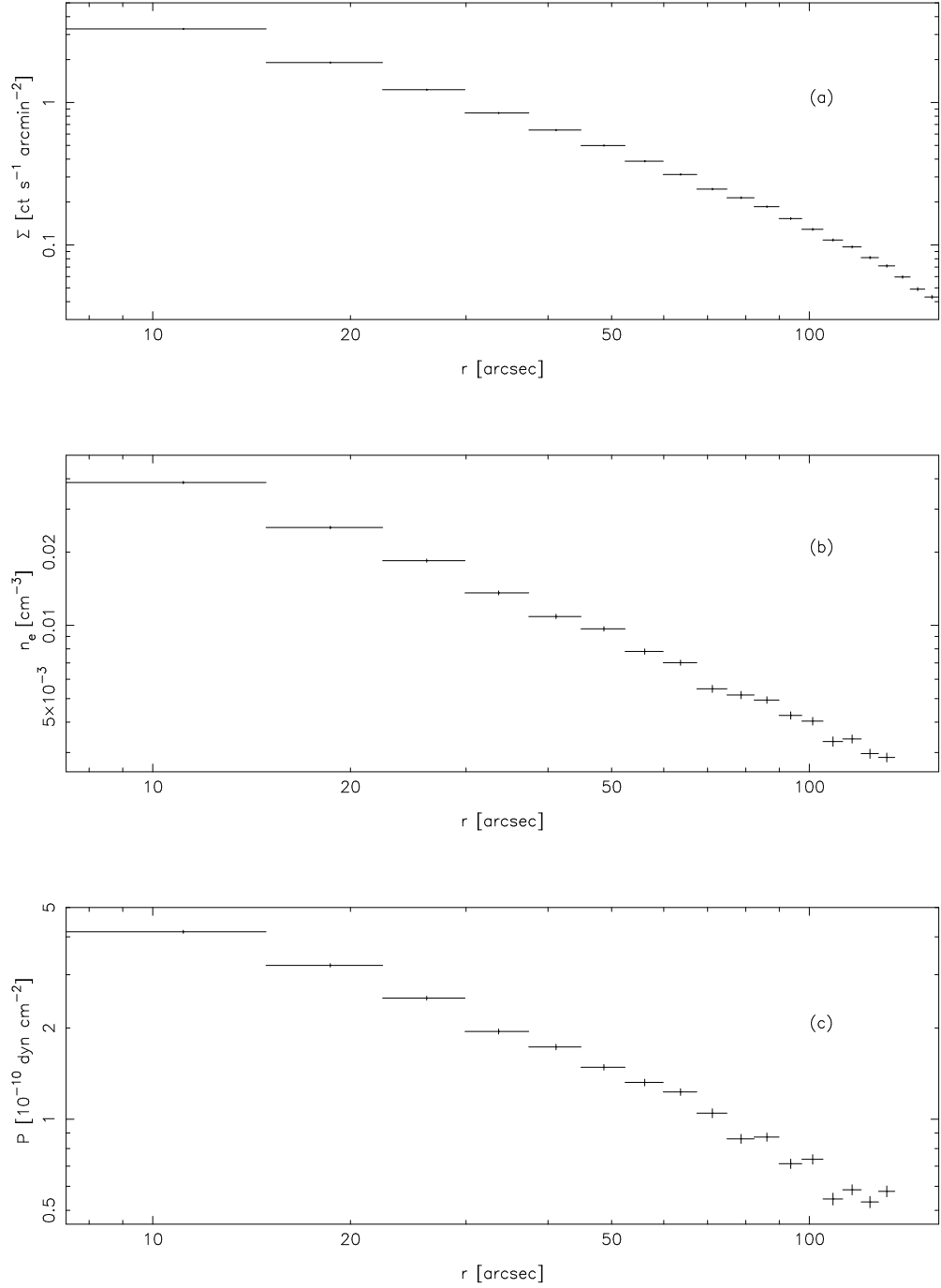}
\caption
{Radial profiles of surface brightness (a), electron density (b), and
 pressure (c), respectively.
 \label{fig:radsfpden}}
\end{figure}

\begin{figure}
\epsscale{0.8}
\plotone{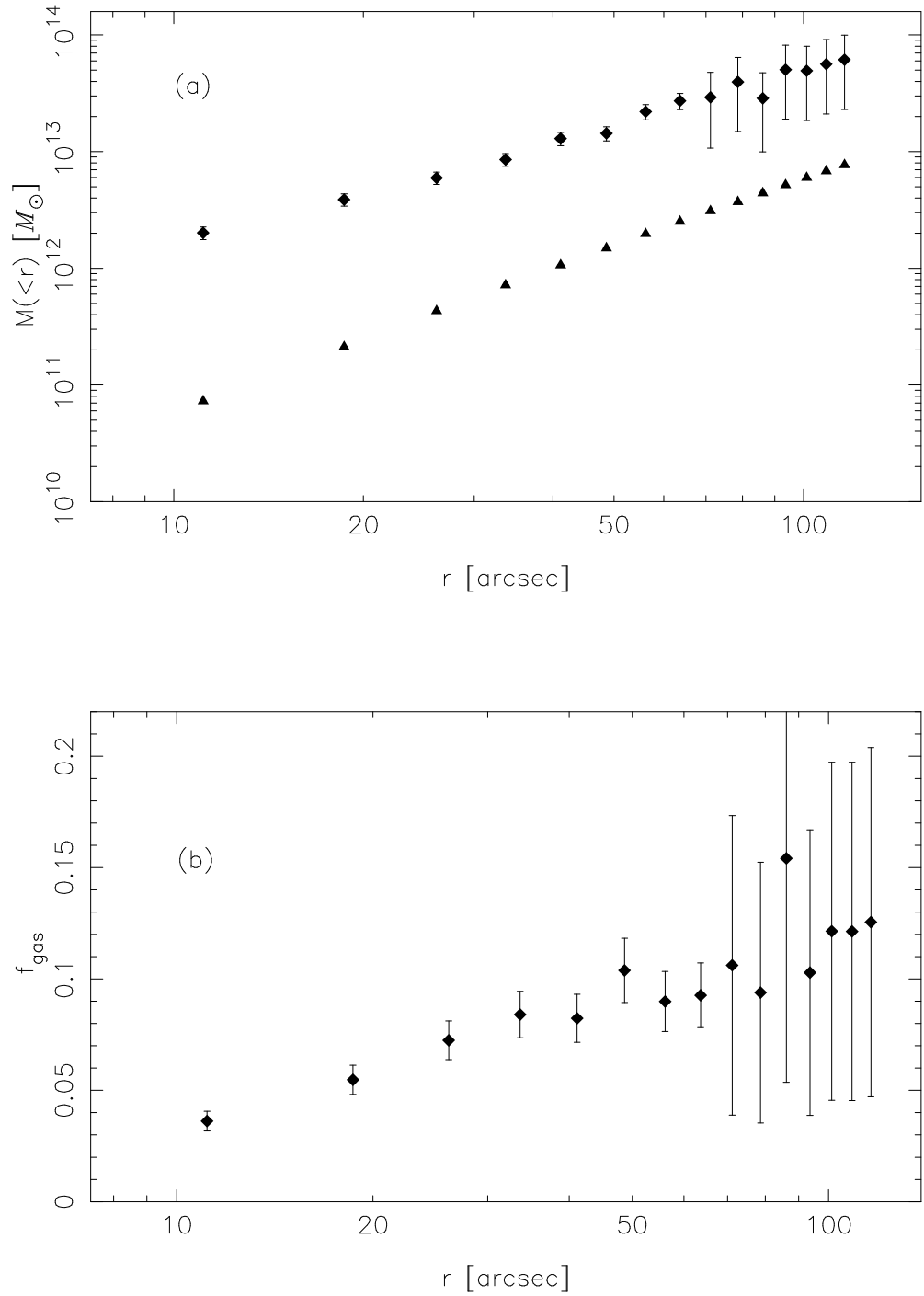}
\caption
{(a) Integrated total gravitational mass (diamonds) and gas mass 
 (triangles).
 (b) Radial profile of gas mass fraction.
  The gas mass fraction increases from $3\%$ at $10\arcsec$ up to $10\%$ at 
  $100\arcsec$.
 \label{fig:radmrmg}}
\end{figure}

We found that the central 
radio source associated with the central cD galaxy in Abell~3112
is interacting with the surrounding ICM
(Fig.~\ref{fig:center1} and \ref{fig:center2}).
The radio image shows
a central core and two diffuse lobes to the SE and SW of the core.
The central X-ray image has an asymmetric 
structure, and the excess X-ray emission over an elliptical isophotal model
roughly appears to surround two radio lobes.
This configuration is naturally 
explained if the radio lobes have swept up the ICM which was
where the radio lobes are now.
Similar scenarios have been proposed in other cooling 
flow clusters with radio sources
(e.g., McNamara et al.\ 2000; Fabian et al.\ 2000; Blanton et al.\ 2001 ). 
In addition, there is excess X-ray emission coincident
with part of the SW radio lobe.
It is possible that the excess X-ray emission is due to
the cool gas that was originally near the very center;
this cool gas might have been entrained by hot buoyant bubbles.
This scenario was originally proposed
to explain the X-ray morphology of the central region of M87
(Churazov et al.\ 2001). 
Indeed, the filamentary structures in the X-ray image of the central region 
of Abell 3112 (Figure \ref{fig:center1}) are similar to those in M87
(Feigelson et al. 1987; B\"{o}hringer et al. 1995).
This model may also apply to Abell 133
(Fujita et al.\ 2002).
In order to determine which model applies to
Abell 3112, temperature and abundance measurements of
the excess X-ray components would be useful because the excess
components are expected to have higher abundances and lower entropy 
in the entrainment scenario.
Unfortunately, the present data have too few counts in these regions to
allow a detailed spectral analysis.
A deeper radio image would also be useful, as would
lower frequency observations to search for steep spectrum emission.
Note that the effect of the central radio source is limited to
a region very close ($r \sim 10\arcsec$) to the center.
This region is much 
smaller than the cooling flow region itself ($r \sim 100\arcsec$). Outside the
interacting region, there is no evidence of the interaction such as
X-ray cavities, instead the X-ray image is quite symmetric and smooth. 
Therefore, the central radio source does not directly affect 
the structure of the cooling flow outside of the very central region, 
at least at present.
Note that we only have high frequency radio data. Low frequency radio
observations might give another picture. In the case of M87 in the Virgo
cluster, for instance, 327 MHz radio data show an morphology which fills out
a large portion of the central region of the cluster 
(Owen, Eilek, \& Kassim 2000), 
although similar structures cannot be seen in higher frequency data.

\begin{figure}
\epsscale{0.8}
\plotone{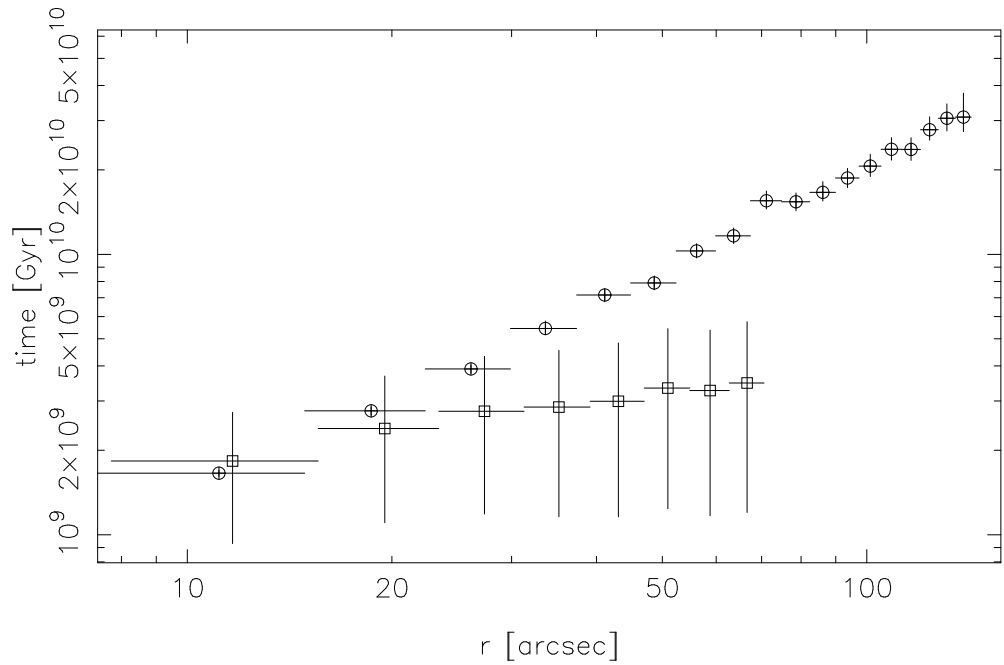}
\caption
{Radial profiles of isobaric cooling timescales (circles) and thermal 
conduction timescales (squares). At the innermost three bins, cooling 
and conduction timescales are comparable with each other. 
Outside this region, conduction timescales are clearly shorter than 
cooling timescales.  
 \label{fig:radtctcond}}
\end{figure}

\begin{figure}
\rotatebox{270}{
    \epsscale{0.7}
    \plotone{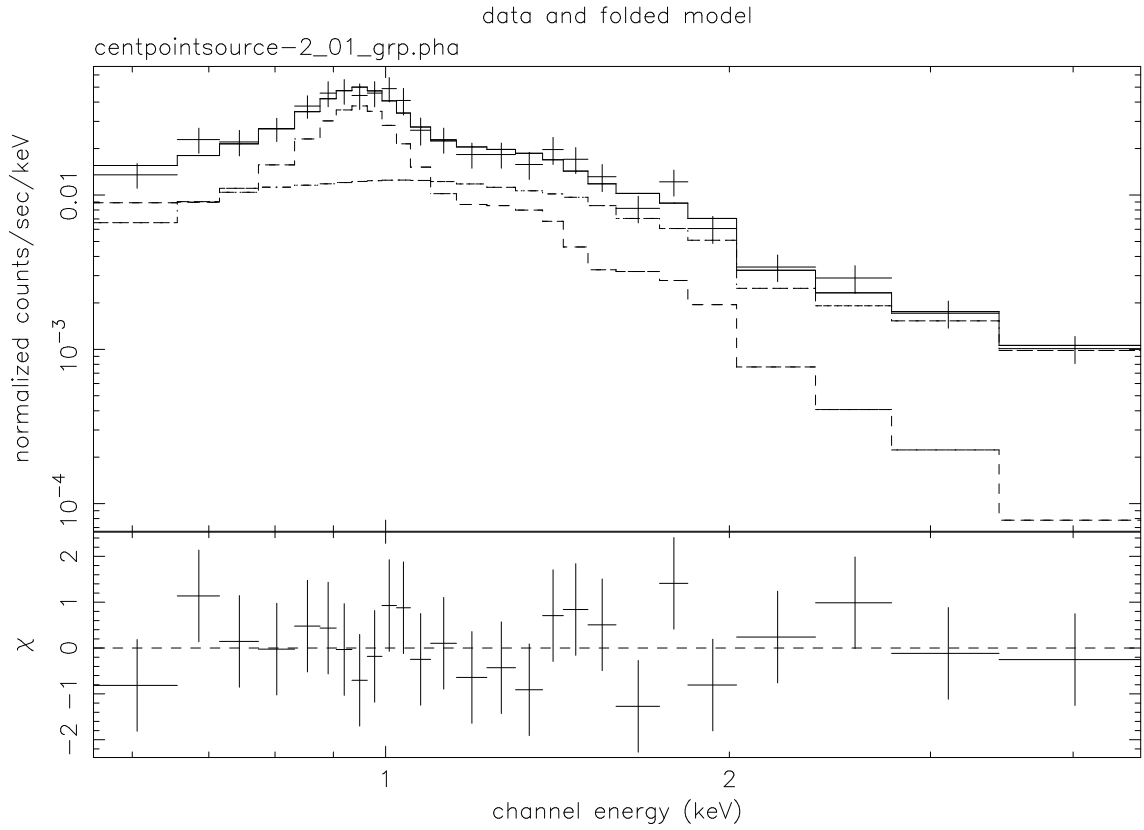}
}
\caption
{Spectrum of the central point source. 
 The spectrum is fit by a 1.26 keV thermal plasma plus 
 a power-law component whose photon index is 1.9.
 The power-law component has an intrinsic extra absorption column of
 $\sim 2 \times 10^{21}$ cm$^{-2}$. This is consistent with the central source
 being a strongly absorbed AGN.
 \label{fig:cpssp}}
\end{figure}

Radiative cooling produces a temperature decrease towards the center of
the cluster.
On the other hand, thermal conduction would reduce this temperature gradient.
If we adopt the standard Spitzer conductivity,
conduction would be expected to be very effective and to have eliminated
the central temperature gradient
(Figure \ref{fig:radtctcond}). 
However, the observations show that the temperature does decrease significantly
at the center of the cluster.
This implies that thermal conduction is significantly suppressed below the
Spitzer value.

Although we did not find a large amount of locally cooling multiphase 
gas, 
the existence of a large-scale temperature gradient indicates that the ICM is
cooling in a global scale. 
The mass deposition rate of Abell 3112 determined from our spectroscopy 
of the total spectrum is comparable to that determined previously based 
on {\it ROSAT} and {\it EXOSAT} image analyses
(Edge, Stewart, \& Fabian 1992; Allen \& Fabian 1997; Peres et al.\ 1998).
However, the temperature range of the cooled gas
is limited to $\geq 2$ keV. In addition, emission measure distribution
about temperature is probably different than that expected from the
standard cooling flow model. 
This suggests that some other heating mechanism may affect the cooling gas.
One probable solution is heat conduction which is reduced from the
Spitzer value by some physical mechanism but still is energetically
important
(e.g., Hattori \& Umetsu 2000; Malyshkin \& Kulsrud 2001). 
The relativistic electrons from the central radio source might also supply the
required heating
(e.g., Fabian et al.\ 2000; McNamara et al.\ 2000; 
David et al.\ 2001; Blanton et al.\ 2001).
However, it is unlikely that the radio-emitting electrons are heating all
of the cooling gas in Abell 3112 at present,
because the radio lobes are so small that they do not directly
affect the cooling flow on large scales.
In Abell~3112, there is no evidence for ghost bubbles which might indicate
a history of past radio source interactions
(McNamara et al.\ 2001).
Another possible form of radio source heating would be
high energy protons from the central AGN.
Since the minimum energy density in relativistic electrons is too
small to support radio bubbles against the surrounding hot gas
(e.g., Blanton et al.\ 2001), most of the energy in the radio source may
be in the form of relativistic protons.
The accelerated protons could diffuse and heat the ICM through Coulomb 
interactions (Rephaeli \& Silk 1995; Inoue \& Sasaki 2001). 
Although they do not emit any observable radio radiation,
protons would produce 
several hundred MeV $\gamma$-rays through the decay of neutral
pions produced in collisions with thermal protons.
Future $\gamma$-ray observations will test this hypothesis.

\acknowledgments
We would like to thank
H. B\"{o}hringer,
J. C. Kempner,
and
Y. Fujita
for very useful comments.
Support for this work was provided by the National Aeronautics and Space
Administration through {\it Chandra}\/ Award Number
GO1-2133X
issued by the {\it Chandra}\/ X-ray Observatory Center,
which is operated by the Smithsonian Astrophysical Observatory for and on
behalf of NASA under contract NAS8-39073.
M.~T.\ was supported in part by a Grant-in-Aid from the Ministry of 
Education, Science, Sports, and Culture of Japan (13440061).
Support for E. L. B. was provided by NASA through the {\it Chandra}
Fellowship
Program, grant award number PF1-20017, under NASA contract number
NAS8-39073.

\clearpage

\clearpage

\clearpage

\clearpage

\clearpage

\clearpage

\clearpage

\clearpage

\clearpage

\clearpage

\clearpage
 \begin{table}
  \caption{Cooling Flow Spectra in Each Annulus with Variable Absorption}
  \label{tab:cflow1}
  \begin{center}
   \begin{tabular}{cccccc}
    \hline \hline \\
 $r$   & $kT_{\rm High}$   & $Z$     & $N_{\rm H}$   &  $\dot{M}$   &    $\chi^2/{\rm dof}$ \\
(arcsec)& (keV)            & ($Z_{\odot}$)    & ($10^{20}$ cm$^{-2}$)& ($M_{\odot}$ yr$^{-1}$) &  
   \\
    \hline
$0-20$   & $3.58^{+0.16}_{-0.14}$ & $1.16^{+0.16}_{-0.14}$ & $5.20^{+0.86}_{-0.82}$ 
             & $13.59^{+8.32}_{-8.09}$     &  284.5/200      \\
$20-39$  & $4.65^{+0.47}_{-0.22}$ & $0.79^{+0.15}_{-0.11}$ & $4.42^{+0.80}_{-0.76}$ 
             & $8.68^{+7.14}_{-8.40}$     &  285.3/233      \\
$39-59$  & $5.60^{+0.61}_{-0.45}$ & $0.55^{+0.13}_{-0.12}$ & $4.26^{+0.93}_{-0.89}$ 
             & $13.82^{+8.67}_{-8.34}$     &  226.2/216      \\
$59-79$  & $5.70^{+0.69}_{-0.39}$ & $0.45^{+0.14}_{-0.12}$ & $3.36^{+1.00}_{-0.86}$ 
             & $0.00^{+7.31}_{-0.00}$     &  234.3/205      \\
$79-98$  & $5.74^{+0.90}_{-0.63}$ & $0.38^{+0.15}_{-0.14}$ & $4.32^{+1.21}_{-1.13}$ 
             & $8.37^{+7.59}_{-7.52}$     &  222.9/195      \\
$98-118$ & $5.69^{+0.71}_{-0.50}$ & $0.40^{+0.16}_{-0.16}$ & $3.11^{+1.16}_{-1.04}$ 
             & $0.00^{+4.68}_{-0.00}$     &  216.8/188      \\
$118-138$ & $6.22^{+0.86}_{-0.66}$ & $0.50^{+0.23}_{-0.21}$ & $2.65^{+0.63}_{-1.14}$ 
             & $0.00^{+2.95}_{-0.00}$     &  202.5/181      \\
$138-157$ & $5.70^{+1.36}_{-0.65}$ & $0.47^{+0.24}_{-0.21}$ & $3.49^{+1.64}_{-1.40}$ 
             & $0.15^{+5.41}_{-0.15}$     &  204.5/175      \\
    \hline
   \end{tabular}
  \end{center}
 \end{table}%

\clearpage
 \begin{table}
  \caption{Cooling Flow Spectra in Each Annulus with Fixed Absorption}
  \label{tab:cflow2}
  \begin{center}
   \begin{tabular}{cccccc}
    \hline \hline \\
 $r$   & $kT_{\rm High}$   & $Z$     & $N_{\rm H}$   &  $\dot{M}$   &    $\chi^2/{\rm dof}$ \\
(arcsec)& (keV)            & ($Z_{\odot}$)    & ($10^{20}$ cm$^{-2}$)& ($M_{\odot}$ yr$^{-1}$) &  
   \\
    \hline
$0-20$   & $3.63^{+0.24}_{-0.11}$ & $1.27^{+0.16}_{-0.14}$ & ($2.51$)             
             &  $0.55^{+7.39}_{-0.55}$  & 315.6/201     \\
$20-39$  & $4.87^{+0.32}_{-0.24}$ & $0.88^{+0.13}_{-0.12}$ & ($2.51$)             
             &  $0.80^{+6.64}_{-0.80}$  & 302.5/234     \\
$39-59$  & $5.65^{+0.64}_{-0.42}$ & $0.60^{+0.13}_{-0.12}$ & ($2.51$)             
             &  $5.32^{+6.529}_{-5.32}$  & 237.1/217     \\
$59-79$  & $5.92^{+0.63}_{-0.31}$ & $0.46^{+0.15}_{-0.12}$ & ($2.51$)             
             &  $0.00^{+4.13}_{-0.00}$  & 236.9/206     \\
$79-98$  & $5.81^{+0.92}_{-0.50}$ & $0.43^{+0.15}_{-0.14}$ & ($2.51$)             
             &  $2.20^{+6.06}_{-2.20}$  & 230.1/196     \\
$98-118$ & $5.84^{+0.67}_{-0.39}$ & $0.41^{+0.17}_{-0.15}$ & ($2.51$)             
             &  $0.00^{+3.08}_{-0.00}$  & 217.7/189     \\
$118-138$ & $6.26^{+0.80}_{-0.54}$ & $0.49^{+0.25}_{-0.18}$ & ($2.51$)             
             &  $0.00^{+2.56}_{-0.00}$  & 202.5/182     \\
$138-157$ & $5.94^{+1.25}_{-0.50}$ & $0.50^{+0.26}_{-0.21}$ & ($2.51$)             
             &  $0.00^{+3.44}_{-0.00}$  & 205.85/176     \\
    \hline
   \end{tabular}
  \end{center}
 \end{table}%

\clearpage
 \begin{table}
  \caption{Total Spectrum ($r<157\arcsec$)}
  \label{tab:totspec}
  \begin{center}
   \begin{tabular}{cccccc}
    \hline \hline \\
 $kT_{\rm High}$ & $kT_{\rm Low}$ & $Z$     & $N_{\rm H}$   &  $\dot{M}$   &    $\chi^2/{\rm dof}$ \\
           (keV) & (keV)          & ($Z_{\odot}$)    & ($10^{20}$ cm$^{-2}$)& ($M_{\odot}$ yr$^{-1}$) &  \\
   \hline
$7.42^{+0.12}_{-0.84}$ & $1.86^{+0.46}_{-0.22}$ & $0.58^{+0.05}_{-0.05}$ & $4.01^{+0.32}_{-0.31}$ 
                       & $609^{+51}_{-133}$ & $699.0/474$  \\
$9.07^{+0.50}_{-1.52}$ & $2.02^{+0.31}_{-0.23}$ & $0.68^{+0.06}_{-0.05}$ & $(2.51)$ 
                       & $544^{+34}_{-67}$ & $758.7/475$  \\
    \hline
   \end{tabular}
  \end{center}
Note. -- The upper and lower rows show the results with absorption column density 
        allowed to vary and fixed to the Galactic value, respectively.
 \end{table}%

\clearpage
 \begin{table}
  \caption{Spectrum of the Central Point Source}
  \label{tab:centps}
  \begin{center}
   \begin{tabular}{cccccc}
    \hline \hline \\
  $kT$   & $Z$           & $N_{\rm H}$            &  $\Gamma$   &
$\Delta N_{\rm H}$          & $\chi^2/{\rm dof}$ \\
 (keV)              & ($Z_{\odot}$) & ($10^{20}$ cm$^{-2}$)&              & ($10^{20}$ cm$^{-2}$)&     \\
    \hline
  $1.26^{+0.15}_{-0.17}$ & $(0.77)$ & $(6.40)$ & $1.89^{+0.30}_{-0.38}$      &$18.8^{+18.1}_{13.6}$    & 12.9/21      \\
  $1.26^{+0.14}_{-0.17}$ & $(1.27)$ & $(2.51)$ & $2.00^{+0.26}_{-0.18}$      &$25.3^{+14.8}_{11.4}$    & 12.9/21    \\
    \hline
   \end{tabular}
  \end{center}
Note. -- The abundance and $N_{\rm H}$ of the MEKAL component were derived 
         from the fit of the region surrounding the point source.
         Upper and lower rows show the results when the $N_{\rm H}$ is 
         allowed to vary or fixed to the Galactic value, respectively.

 \end{table}%

\end{document}